\documentclass[onecolumn,letterpaper,11pt]{quantumarticle}
\pdfoutput=1
\usepackage[utf8]{inputenc}
\usepackage{graphicx}

\begin{document}

\title{The solution to the problem of time in quantum gravity also solves the time of arrival problem in quantum mechanics}

\author{Rodolfo Gambini}
\affiliation{
Instituto de F\'{\i}sica, Facultad de Ciencias, Igu\'a 4225, esq. Mataojo,
11400 Montevideo, Uruguay.}
\author{Jorge Pullin}
\affiliation{
Department of Physics and Astronomy, Louisiana State University,
Baton Rouge, LA 70803-4001, USA.}

\date{February 15th 2022}

\begin{abstract}
         We introduced with coauthors some years ago a solution to the problem of time in quantum gravity which consists in formulating  the quantum theory in terms of real clocks. It combines Page and Wootters' relational proposal with Rovelli's evolving constants of the motion. Time is associated with an operator and not a classical parameter. We show here that this construction provides a natural solution to the time of arrival problem in quantum mechanics and leads to a well defined time-energy uncertainty relation for the clocks.
\end{abstract}

\maketitle
\section{Introduction}
In ordinary quantum mechanics there is no operator associated with time, which is a classical parameter. This has led to the ``time of arrival'' problem: one can make probabilistic predictions about where a particle will be detected at a certain time, but not about when a particle will be detected at a certain position. 
This problem has been addressed through various approaches in the past and there does not appear to be a consensus answer for it. See \cite{review} for reviews of previous work.  We argue that a previous proposal to address the problem of time in constrained systems like quantum gravity naturally leads to a solution of the time of arrival problem. It also leads to a well defined time-energy uncertainty relations for the clocks. Our proposal has elements in common with that of Maccone and Sacha \cite{masa} but they do not consider evolving constants of the motion as their focus is not on totally constrained systems like general relativity. In particular, their analysis does not include the computation of the time of arrival measured by real clocks that evolve with a Hamiltonian that is bounded-below. 

We introduced with coauthors a few years ago \cite{Pedagogical,Torterolo}, a solution to the problem of time in generally covariant systems like quantum general relativity. In such systems the Hamiltonian is a linear combination of the constraints, and therefore  all Dirac observables are constants of the motion, so time evolution has to be introduced in a different way that in traditional classical mechanics. The solution is based in considering the conditional probabilities that Page and Wootters \cite{PaWo} had proposed  for solving the problem, but between parameterized Dirac observables (evolving constants of the motion). This provides observable quantities that evolve in time and are well defined in the physical space of solutions of the constraints. It eliminates the objections that Kucha\v{r} \cite{kuchar} had made to the Page Wotters construction and leads to correct propagators in model systems. Physical time arises from observing a (quantum) clock. Our proposal has also been shown to be compatible with the approach in terms of positive operator-valued measures of H\"ohn, Smith and Lock \cite{hoehn}. The idea is to pick a parameterized Dirac Observable one wishes to study $O(t)$ and pick another one that will play the role of a clock $T(t)$. In these expressions $t$ is the parameter the Dirac observables depend on. Parametized Dirac observables were introduced by \cite{evolving} for the description of the evolution. Generically they are Dirac observables that depend on one or several parameters such that when certain variables in the kinematical space of the constrained system take the value of these parameters the observable reproduce the value of the other variables in the kinematical space. In field theories like general relativity the parameters may be functions \cite{GOP2014}. One then computes the conditional probabilities of the observable taking a value within an interval of width $2\Delta O$ around a value $O_0$ at a time within an interval of width $2\Delta T$ around a value $T_0$,  
\begin{equation}
    {\rm P}\left(O\in [O_0\pm \Delta O]\vert T\in [T_0\pm \Delta T]\right) =\lim_{\tau\to \infty} \frac{\int_{-\tau}^\tau dt 
    {\rm Tr}\left(P_{O_0}\left(t\right) P_{T_0}\left(t\right) \rho P_{T_0}\left(t\right)\right)}{\int_{-\tau}^\tau dt {\rm Tr}\left(P_{T_0}\left(t\right)\rho\right)}.\label{1}
\end{equation}
In this expression $\rho$ is the density matrix of the system, $P_{O_0}$ is the projector on the eigenspace associated with the eigenvalue $O_0$ of $O$ and similarly for $P_{T_0}$. The classical parameter $t$ is integrated over because it is not an observable quantity, its determination would require the observation of a non observable kinematical variable,  and we do not need knowledge of it to compute the probability. Working with evolving constants of the motion is equivalent to working  in a Heisenberg like representation where operators depend on the parameter $t$ and the density matrices do not. Like in the case of H\"ohn, Smith and Lock \cite{hoehn} we assume that the evolution in the ideal parameter $t$ is at a constant rate (otherwise there would be a non-trivial measure in the integral in $t$). The above expression is for operators with continuous spectrum (hence the intervals in the values), similar expressions can be introduced for operators with discrete spectrum.

We have been able to show that if one assumes that the density matrix of the systems is a direct product of that of the system under study and that of the clock $\rho=\rho_{\rm sys} \otimes \rho_{\rm cl}$ and so are the evolution operators $U=U_{\rm sys}\otimes U_{\rm cl}$ (the system and the clock do not interact) and that the clock behaves almost like a perfect clock, the resulting density matrix of the system obeys a modified Schr\"odinger equation. To be a bit more precise, if one defines the (physically unobservable as it depends on the parameter $t$) probability
that the parameter takes the value $t$ when one observes the clock taking a value $T_0$,

\begin{equation}
{\rm P}_t\left(T_0\right)=\frac
{{\rm Tr}\left(P_{T_0}\left(t\right)U_{\rm cl}\left(t\right) \rho_{\rm cl} U_{\rm cl}\left(t\right)^\dagger\right)}
{\int_{-\infty}^\infty dt {\rm Tr}\left(P_{T_0}\left(t\right) \rho_{\rm cl}\right)}
\end{equation}
for an ideal clock it would be a Dirac delta. In terms of this probability one can label quantum states in terms of the readings of the real clock as,
\begin{equation}
    \rho(T)=\int_{-\infty}^\infty dt U_{\rm sys}\left(t\right)\rho_{\rm sys}U^\dagger\left(t\right) {\rm P}_{t}\left(T\right).
\end{equation}
We see that if ${\rm P}_t\left(T\right)$ were a Dirac delta one would recover the usual evolution in quantum mechanics. Otherwise, the theory will lose unitarity due to the real clock $T$ not being able to track precisely the parameter time $t$ in terms of which evolution is unitary.  A detailed analysis of the master equation that describe the evolution with real clocks may be found in \cite{Pedagogical}.

We will apply this notion of time to the understanding of the unresolved problems that affect the the solutions of the time of arrival problem in non-relativistic quantum mechanics proposed up to now. We start by analyzing the evolution of a quantum system in terms of another physical system that we take as a clock. 
To illustrate these ideas, let us start with a situation in which one has an ideal clock. We represent it as a particle whose Hamiltonian is proportional to the momentum $p_1$. We will later consider more realistic clocks whose Hamiltonians are bounded below. The system  under study will be a particle with a usual non-relativistic Hamiltonian with a potential, that is,
\begin{equation}
H_1=c\, p_1,\qquad H_2=\frac{p_2^2}{2m_2} +V\left(x_2\right),
\end{equation}
so $H_1$ is the Hamiltonian we choose for  the clock and $H_2$ that of the system under study.
We turn this into a constrained system by the standard procedure of parameterization. That is, we introduce a conjugate momentum $p_0$ to the non-relativistic evolution variable $x_0$. So our canonical pairs will be $x_0,p_0$, $x_1, p_1$ and $x_2, p_2$. The system will therefore have a constraint,
\begin{equation}
    \phi=p_0+H_1/c+H_2/c,
\end{equation}
with $c$ the speed of light.

We perform a usual canonical quantization promoting the canonical pairs to operators acting on a suitable kinematical Hilbert space of functions of $x_0, x_1, x_2$ or their conjugate momenta.

We start by identifying the Dirac observables of the system, which have vanishing commutators with the constraint $\left[\hat{x}_i^{\rm Dirac},\phi\right]=0$,
\begin{eqnarray}
\hat{x}_1^{\rm Dirac}&=&\hat{x}_1-\hat{x}_0,\\
\hat{x}_2^{\rm Dirac}&=&\exp\left(-i\frac{\hat{H}_2}{c} \hat{x}_0\right) \hat{x}_2 \exp\left(i \frac{\hat{H}_2}{c} \hat{x}_0\right).
\end{eqnarray}

We can also consider parameterized Dirac observables (evolving constants of the motion),
\begin{eqnarray}
\hat{x}_1\left(t\right)&=& \hat{x}_1^{\rm Dirac}+ c t,\\
\hat{x}_2\left(t\right)&=&
\exp\left(i \hat{H}_2 t\right) \hat{x}_2^{\rm Dirac} \exp\left(-i \hat{H}_2 t\right),
\end{eqnarray}
where $t$ is a classical parameter, whose classical counterparts $x_1(t)$ and $x_2(t)$ satisfy that $x_1(t=x_0/c)=x_1$ and $x_2(t=x_0/c)=x_2$ and we choose $\hbar=1$. These operators describe the position of particles $1$ and $2$. Notice that $\hat{x}_1$ and $\hat{x}_2$ are not Dirac observables, but $\hat{x}_1\left(t\right)$ and $\hat{x}_2\left(t\right)$ are, they have vanishing commutators with the constraint. They are what Rovelli calls {\em evolving constants of the motion}. We will use the position of particle $1$ as a clock.

The physical space of states for this system can be constructed straightforwardly. If $\hat{H}_2 \psi_E\left(p_2\right)=E \Psi_E\left(p_2\right)$ are the eigenfunctions of $\hat{H}_2$ with eigenvalue $E$, we have that a physical state for the complete system (a state that is annihilated by the constraint acting as an operator) is given by,
\begin{equation}
    \Psi_{\rm Ph}\left(p_1,p_2,p_0\right)=
    \int dE \delta\left(p_0+p_1+E/c\right)f\left(p_1\right) \Psi_E\left(p_2\right) C_E,
\end{equation}
where $C_E$ are the complex coefficients of the expansion of the state in the energy basis.
where we have assumed  that $\hat{H}_2$ has a continuous spectrum and $1$ and $2$ are independent. A similar construction can be made if the spectrum is discrete, the formulas vary slightly. 

The inner product in the physical Hilbert space ${\cal H}$ can be constructed using well known techniques for constrained systems \cite{AsMoetal, GaPo},
\begin{eqnarray}
\langle \Psi\vert \Psi'\rangle_{\rm Ph}&=&
\int dp_1 dp_2 dE dE' f^*\left(p_1\right) \psi^*_E\left(p_2\right) C^*_E C'_{E'} f'\left(p_1\right)\Psi_{E'}\left(p_2\right),\nonumber\\
&=&\int dp_1 dE f^*\left(p_1\right) f'\left(p_1\right) C^*_E C'_E,
\end{eqnarray}
where we have taken into account that the eigenfunctions of $H_2$ are orthogonal for different eigenvalues $E$. Using the expression for the conditional probabilities (\ref{1}) and   working with $\hat{x}_1\left(t\right)$ and $\hat{x}_2\left(t\right)$ we have,
\begin{equation}
{\rm P}\left(x_2\in [y\pm \Delta y/2]\vert x_1=x\right)=
\lim_{\tau\to\infty}
\frac
{\frac{1}{2\tau}\int_{-\tau}^\tau dt {\rm Tr}\left(P\left(y\right) \rho_2\left(t\right)\right) 
{\rm Tr}\left(P\left(x\right)\rho_1\left(t\right)\right)}
{\frac{1}{2\tau}
\int_{-\tau}^\tau dt {\rm Tr}\left(P\left(x\right)\rho_1\left(t\right)\right)
},\label{11}
\end{equation}
where we have assumed that ${\rm Tr}\left(\rho_1\right)={\rm Tr}\left(\rho_2\right)=1$ and taken into account that $P_x(t)=U^\dagger(t) P_x(0) U(t)$ and that $\rho_{1,2}(t)=U(t)\rho_{1,2} U^\dagger(t)$. Here $\rho_1$ and $\rho_2$ are the density matrices of particles $1$ and $2$ respectively. In this example $\rho_1$ would play the role of $\rho_{\rm cl}$ and $\rho_2$ that of $\rho_{\rm sys}$ and as we discussed in the introduction they are assumed not to interact with each other. As before $P(y)$ is the projector on the eigenspace associated with the eigenvalues in the interval around $y$, i.e. $P(y)=\int_{y-\Delta y/2}^{y+\Delta y/2} dz \vert z\rangle\langle z\vert$, and $P(x)=|x><x|$. We do this because normally the description of probabilities in quantum mechanics is for a given time and not for a time interval. 
Notice that here $t$ is the parameter previously introduced, that as we have discussed is not observable. All the previous relations were derived without any additional assumption about the nature of $t$.  We shall call $t$ the ideal time, the explicit connection with the usual definition of time can be obtained by imposing a gauge fixing $x0=c t$. If we did that we would notice that the operators ${\hat x}_1(t)$ and ${\hat x}_2(t)$ are the usual ones for those systems in the Heisenberg representation. That allows to consider the parameter $t$ as the usual time in quantum mechanics, that we consider not accessible since we measure times with physical clocks\footnote{We live in a quantum mechanical generally covariant Universe where all observable quantities are constants of the motion represented by Dirac observables}. We omit the hats on the projectors and evolution operators to keep in line with notation in our previous papers. We see that in the expression the ideal time $t$ is integrated over and therefore we do not need to know its value in computing the probability, as we mentioned before.

Let us analyze the behavior of the clock when it is perfectly synchronized with the ideal time $t$. That is, when the system $1$ behaves like an ideal clock. In that case we would have that 
$\Psi\left(x_1,t\right)=\Psi\left(x_1-c t\right)$ since,
\begin{eqnarray}
\hat{x}_1\left(t\right) \vert x,t\rangle&=& x \vert x,t\rangle=\left(x_1^{\rm Dirac}+c t\right) \vert x,t\rangle,\\
\hat{x_1}^{\rm Dirac} \vert x,t\rangle&=& \left(x-c t\right)\vert x,t\rangle.
\label{15}
\end{eqnarray}
Since $p_1$ is the Hamiltonian of the first particle we have that $\langle x, t\vert=\langle x\vert \exp(-i \hat{p}_1 t)$. Therefore $\psi(x,t)=\langle x,t\vert \psi\rangle=\langle x-c t\vert\psi\rangle$.

In order to have a reasonably behaved clock we need a localized wavefunction, for instance a Gaussian, that advances as $t$ grows,
\begin{equation}
    \Psi_1\left(x,t\right)=\left(\frac{1}{2\pi}\right)^{1/4} 
    \frac
    {\exp\left(-\frac{\left(x-c t\right)^2}{4{\sigma_x}^2}\right)}
    {{\sigma_x}^{1/2}},
\end{equation}
and 
\begin{equation}
    {\rm Tr}\left(P\left(x\right) \rho_1\left(t\right)\right)=
    \frac{1}{\sqrt{2\pi}}
    \frac{\exp\left(-\frac{\left(x-c t\right)^2}{2{\sigma_ x}^2}\right)}
    {\sigma_x}.
\end{equation}

The position of system $1$ behaves like an ideal clock when $\sigma_x \to 0$, where one has that,
\begin{equation}
    {\rm Tr}\left(P(x)\rho_1^{\rm ideal}
    \left(t\right)\right)=\delta\left(x-c t\right).
\end{equation}

As a consequence, the denominator of equation (\ref{11}) becomes $1/(2\tau)$ for all $x\in [-\tau,\tau]$ and therefore,
\begin{eqnarray}
{\rm P}\left(x_2\in [y\pm \Delta y/2]\vert x_1=x\right)&=&
\lim_{\tau\to \infty} 
 \int_{-\tau}^\tau dt {\rm Tr}\left(P\left(y\right)\rho_2\left(t\right)\right) {\rm Tr}\left(P\left(x\right)\rho_1\left(t\right)\right)\nonumber\\
&\propto&\lim_{\Delta x \to 0}\frac{1}{\Delta x}{\rm P}\left(x_2\in[y\pm \Delta y/2],x_1\in[x\pm \frac{\Delta x}{2}]\right)\label{18}
 \end{eqnarray}

and the conditional probability coincides (up to a factor) with the simultaneous probability of measuring $x_1$ and $x_2$ at certain instant. 
It should be noted that the conditional probability (\ref{18}), for $\Delta y$ sufficiently small and $\rho_2(t)=|\psi_2(t)><\psi_2(t)|$ is nothing else but 
$|\psi_2(y,t)|^2\Delta y$, that is, the Born rule applied at $t=x/c$. We have therefore shown how the covariant description in terms of evolving Dirac observables allows to recover, when the clock is perfectly correlated with $t$ the usual Born rule.

\section{Time of arrival in terms of conditional probabilities}

The use of real clocks allows to assign observables to the time variable. In the previous example $\hat{x}_1(t)$, the evolving constant of the motion representing the position of particle $1$, plays the role of time. In this section we will define the probabilistic distribution of the time of arrival of a particle at a given point. Different requirements to assign probabilities to the time of arrival have been developed through the years. All of them have questionable aspects and lead to paradoxes. We will see that the solution here proposed not only has a very simple and clear origin but it also solves the paradoxes. We will concentrate in the comparison with the ``standard approach to time of arrival'' advocated by Egusquiza, Muga and Baute \cite{egmuba}  and discussed by several of the authors in \cite{review}, among them the derivations of Allcock, of Kijowski, of Grot, Rovelli and Tate, and of Delgado and Muga are for the special case of free  evolution. They \cite{egmuba} titled their distillation of the various approaches leading to the two equations that appear bellow ``standard” quantum mechanical approach to times of arrival, emphasizing their claim that (2) can be derived ``without in any way distorting the standard framework of quantum mechanics”. However, Leavens  \cite{Le} observed that even for the case of free evolution,  they need to associate the direction of arrival with sign of $k$, which is  an unjustifed assumption that is not a part of conventional quantum mechanics. As we will discuss in detail in section 3, several paradoxical behaviours emerge.

 We will follow the latter in order to derive the expression of the distribution of times of arrival for an ensemble of quantum particles whose initial state is $\psi\left(x,t=0\right)$.

The standard approach starts from the probability distribution of observing the particle arrival at $X$ at time $T$,
\begin{equation}
    \Pi\left(T,X\right)=\Pi_+\left(T,X  \right)
    +\Pi_-\left(T,X\right),
\end{equation}
with,
\begin{equation}
    \Pi_\pm\left(T,X\right)=
    \frac{1}{2\pi m}
    \left\vert
    \int_{-\infty}^\infty dk \Theta\left(\pm k\right) \vert k\vert^{1/2} \exp\left(ikX\right)\phi\left(k,T\right)
    \right\vert^2 \label{eq2}
\end{equation}
where $\Pi_+$ and $\Pi_-$ represent the contributions to $\Pi\left(T,X\right)$ from particles that come from the left and right respectively. $\phi(k,T)$ is the momentum space representation of the quantum state (Leavens \cite{Le} calls it $\phi$, we call it $\psi$ in the rest of this paper) The expression was initially derived for free particles and was recently extended \cite{bauteetal} for particles in an arbitrary potential $V\left(X,T\right)$. Leavens \cite{Le} has analyzed in several concrete examples the counter-intuitive and paradoxical aspects of the standard approach. We will revisit them here using the technique of conditional probabilities and we will see that the issues are resolved. We will leave for a future publication the comparison of these results with those of Bohmian mechanics that also appear to solve some of the observed problems.

We are interested in computing the probability that the clock indicates $x_1=cT$ when the system is at $x_2=y$,
\begin{equation}
    {\rm P}\left(
    x_1\in [cT\pm c\Delta T] \vert
    x_2\in [y\pm \Delta y]\right)=
    \lim_{\tau\to\infty}
    \frac
    {\int_{-\tau}^\tau dt
    {\rm Tr}\left(P_T(t) \rho_1\right)
    {\rm Tr}\left(P_y(t) \rho_2\right)}
    {\int_{-\tau}^\tau dt 
    {\rm Tr}\left(P_y(t)\rho_2\right)}.
    \label{eq3}
\end{equation}

Let us apply the technique to the free particle $V(x_2)=0$. For concreteness, let us consider the evolution of a Gaussian wave packet of initial width $\sigma_y$ prepared at $t=0$ and centered in $y=0$. One can immediately go to the Schr\"odinger picture where projectors are time independent and $\rho_2(t)=\vert \psi(t)\rangle\langle \psi(t)\vert$ with,
\begin{equation}
    \langle y\vert \psi(t)\rangle=
    \frac{\left(2\pi\sigma_y^2\right)^{-1/4}}{\sqrt{1+i\alpha t}}
    \exp\left[
    \frac{-\frac{y^2}{2\sigma_y^2}+i p_0 y -i\frac{p_0^2}{2m}t}
    {1+i\alpha t}
    \right]\Theta(t),
\end{equation}
with $\alpha=1/(2m\sigma_y^2)$, $\hbar=1$ and $\Theta(t)$ a Heaviside function equal to $0$ for $t<0$ and $1$ otherwise. The initial expectation value of the momentum of the particle is $p_0$. We have chosen zero as the instant of the preparation of the state. The probability density corresponding to the state as a function of time is,
\begin{equation}
    {\rm P}(y,t)=
    \frac{\left(2\pi\sigma_y^2\right)^{-1/2}}
    {\sqrt{1+\alpha^2 t^2}}
    \exp\left[
    -\frac{2(y-v_o t)^2}
    {2 \sigma_y^2  \left(1+\alpha^2t^2\right)}
    \right]\Theta(t),\label{eq4}
\end{equation}
with $v_0=p_0/m$ and ${\rm Tr}\left(P_y \rho_2(t)\right)={\rm P}(y,t)\Delta y$.

In the limiting case in which particle $1$ behaves like an ideal clock, one has that ${\rm Tr}\left(P_T\rho_1(t)\right)=\delta(T-t)\Delta T$ and the probability density that the time of arrival to $y$ be $T$ is,
\begin{equation}
    {\rm P}\left( x_1\in [cT\pm \frac{c\Delta T}{2}]\vert x_2 \in [y\pm \frac{\Delta y}{2}]\right)=\frac{{\rm P}(y,T)}{\int_0^\infty dt {\rm P}(y,t)},\label{24}
\end{equation}
with ${\rm P}(y,T=t)$ given by (\ref{eq4}). The expression obtained differs from that of the standard approach \cite{Le}, which is given by  (\ref{eq2}) with,
\begin{equation}
    \phi(k,T)=\left(\frac{2\sigma_y^2}{\pi}\right)^{1/4} \exp\left(-\left(k-p_0\right)^2 \sigma_y^2 -i\frac{k^2}{2m} T\right).
\end{equation}
However, as shown in figure 1 for a choice of parameters given by $p_0=100,  m=50, \sigma_y=0.1$ and therefore $v_0=2$ the qualitative behavior of the time of arrival is very similar, differing in slightly over $1\%$. Recall that we are working in unis such that $\hbar=1$ and distances, velocities and times are in compatible units, for instance $cm$, $cm/s$, $s$.

We therefore see that the approach we advocate reproduces the traditional results for the case without a potential. That case does not involve paradoxes and contradictions, so both approaches can be considered satisfactory. To illustrate the advantages of our approach, we will discuss other situations in the next section.

\begin{figure}
    \centering
    \includegraphics[scale=0.35]{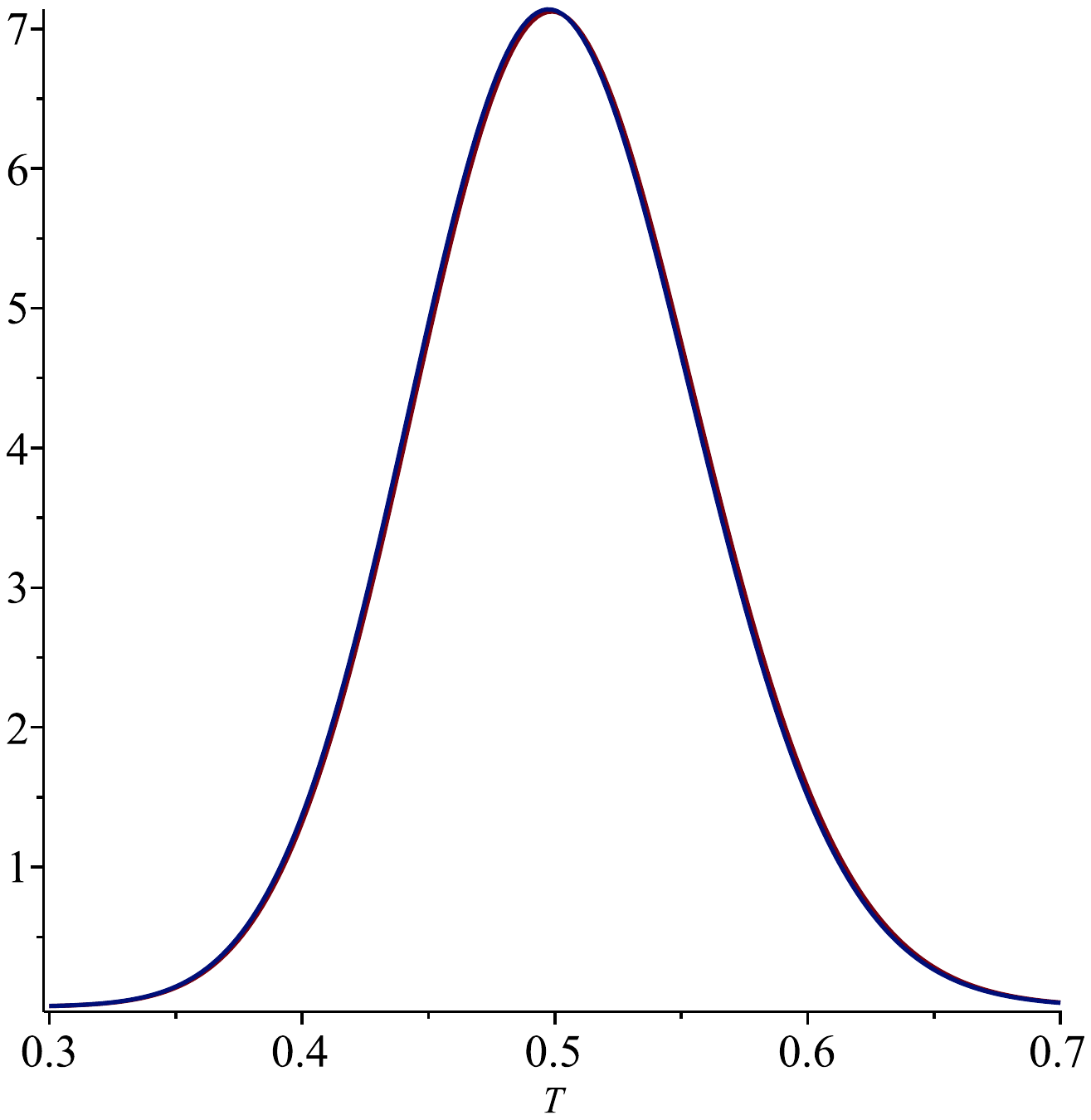}
    \includegraphics[scale=0.35]{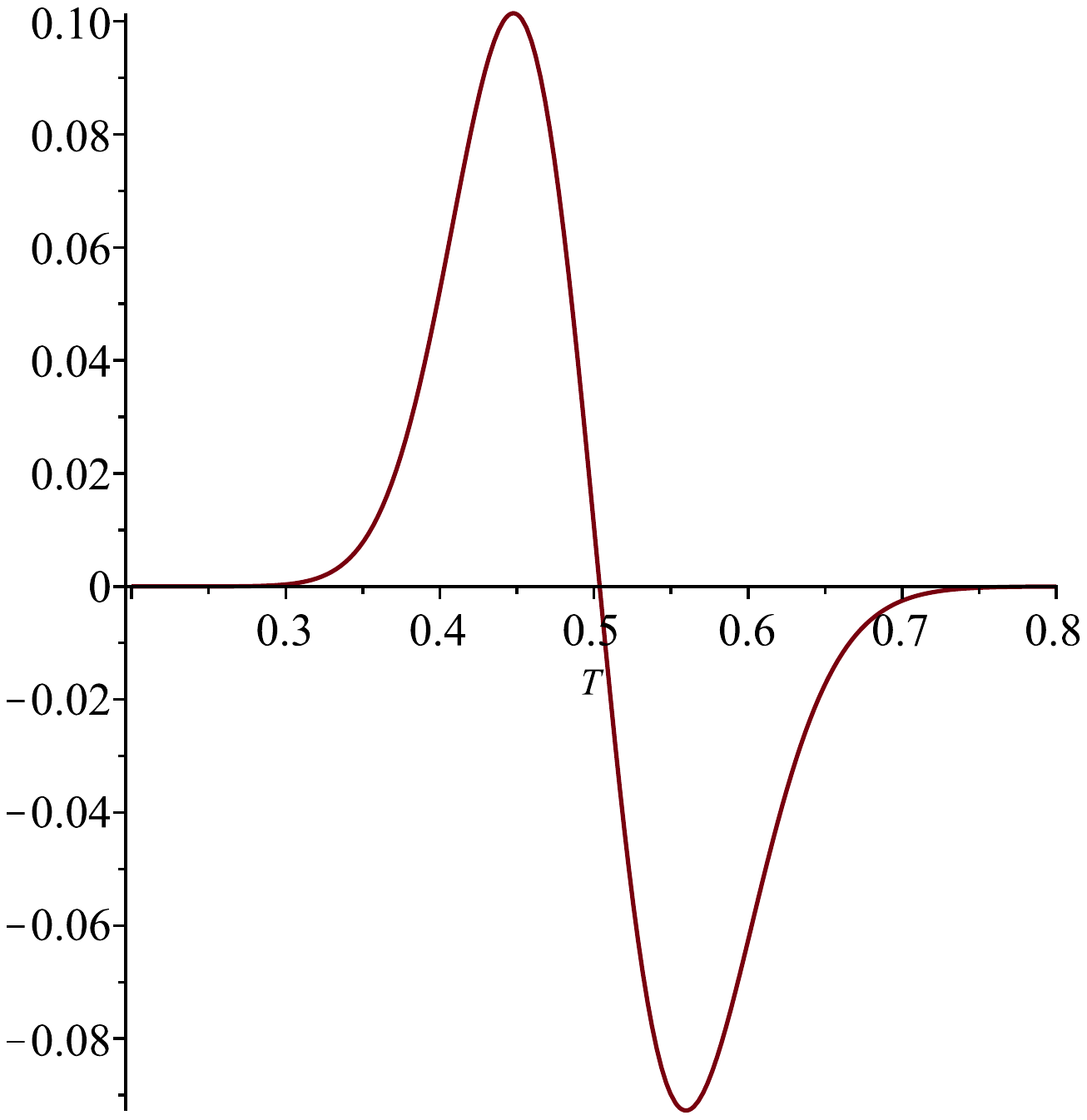}
    \vspace{-4cm}
    \caption{The left pane shows the probabilities of arrival at $y=1$ as a function of time for a particle with velocity $1/2$ computed with the standard approach and with the conditional probability approach we present in this paper, for the example of a Gaussian wave packet discussed in the text.  The right pane shows the difference between them, which amounts to slightly over $1\%$. Conditional probabilities assign an expectation value for $T$ slightly higher than the one resulting from the standard approach. Similar results are obtained for all values of the parameters we tested.}
    \label{fig:my_label}
\end{figure}

\section{Freedom from usual paradoxes}

When the standard approach is applied to an infinite potential barrier, it leads to paradoxical predictions. A good discussion of the problem can be found in Leavens \cite{Le}. He shows that, applied in that case, the standard approach leads to a non-vanishing distribution of arrival times in the forbidden regions where potentials are infinite. The infinite barrier is a particular case of an infinite well when its width tends to infinity. Leavens' analysis also leads to non-trivial arrival times in regions where the probability of finding the particle is strictly zero in the case of the well. As a consequence, one can easily see that the same will happen to the infinite barrier. Figure 1 in that paper show the arrival times in the forbidden region for arbitrary values of the well's width $a$. 

Let us start from equation (\ref{24}) and consider the case of a potential barrier at $y=0$. The particle is in the region $y>0$. Let us study a wavepacket initially centered in $p_0, y_0$, confined to move in the region $y>0$,
\begin{equation}
    \psi_2\left(y,t\right)=\left(\frac{2 \sigma_y^2}{\pi}\right)^{1/4} \int_{-\infty}^\infty \sin\left(p y\right) \Theta(y) \exp\left(-\left(p-p_0\right)^2\sigma_y^2\right) 
    \exp\left(-\frac{p^2 t}{2 m}\right) dp\, \Theta(t),\label{26}
\end{equation}
where we have chosen the origin of the $t$ parameter at the instant of preparation of the state. The conditional probability of observing the clock at $x_1$ in $T$ when it goes through $x_2=y$ is given by,
\begin{equation}
    {\rm P}\left(x_1=T\vert x_2=y\right)=\frac{{\rm P}(y,t=T)}{\int {\rm P}(y,t) dt},
\end{equation}
with,
\begin{eqnarray}
{\rm P}(y,t)&=& \frac{1}{2\sqrt{\frac{\alpha^2 t^2+1}{\alpha^2}}}\nonumber
\left(\sqrt{\frac{\pi m}{\alpha}} \Theta(y)\left[ \exp\left(-\frac{\beta_-}{\rho}\right)
+\exp\left(-\frac{\beta_+}{\rho}\right)\right.\right.\nonumber\\
&&\left.\left.
-\exp\left(\frac{\gamma+\delta}{\rho}\right)
-\exp\left(\frac{\gamma-\delta}{\rho}\right)\right)\right]\Theta(t)
\end{eqnarray}
with $\Theta(y)$ and $\Theta(t)$ Heaviside functions and,
\begin{eqnarray}
\alpha&=&\frac{1}{2 m \sigma_y^2},\\
\beta_\pm&=& \alpha\left(m(y \pm y_0)\pm t p_0\right)^2,\\
\rho&=& m\left(\alpha^2 t^2+1\right),\\
\gamma&=&\left(\left(-y^2-y_0^2\right)m^2-2 m p_0 t y_0-p_0^2t^2\right)\alpha,\\
\delta&=& 2 i mp_0 y-2 i\alpha^2 m^2 t y y_0.
\end{eqnarray}

One can immediately check that for $y<0$ the expression yields $0/0$ and is ill defined. As a consequence, there is no prediction of a time of arrival in the forbidden region. 

Let us analyze in some detail two cases: 1) an initially well localized packet at $y=3$ that propagates towards the barrier with $p/m=v=-4$.
 The time of arrival is shown in figure 2.
\begin{figure}[h]
    \centering
    \includegraphics[scale=0.55]{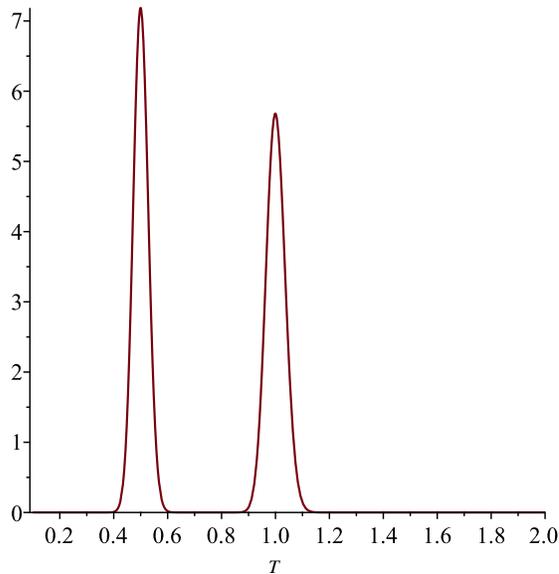}
    \vspace{-6cm}
    \caption{Time of arrival at $y=1$ for a packet prepared at $y=3$ with velocity $v=-4$  in the case of an infinite potential barrier at $y=0$. The two peaks corresponding to the first passage of the particle at time $x=1/2$ and the passage of its reflection at time $x=1$. }
\end{figure}

The probability distribution of measuring $t$ has two peaks, one at $T=0.5$ that represents the direct propagation from $y=3$ to $y=1$ and another at $T=1$ which represents the arrival after the reflection of the packet on the barrier. The packet widens with time so the latter arrival time will have more dispersion in $T=1$. Figure 3 shows the arrival time for a particle that starts at $y=3$ to $y=1$ that moves with speed $-0.1$. Due to lower speed, the widening of the packet generates interference between the wave that travels towards the barrier and the reflected one. A classical particle would have arrival times $T=20$ and $T=40$. The dispersion allows to find the particle in $y=1$ for large times. 
\begin{figure}[h]
    \centering
    \includegraphics[scale=0.45]{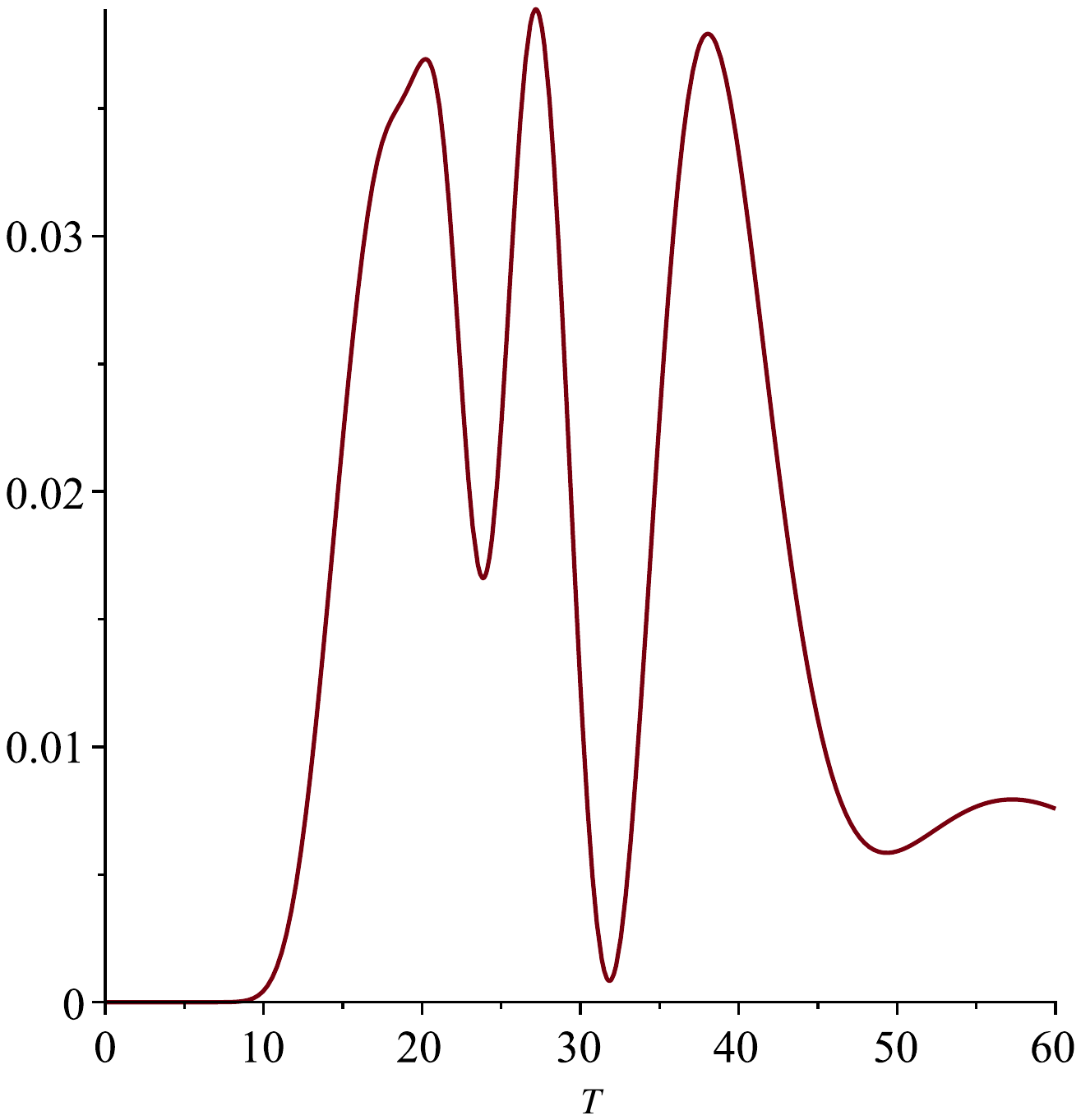}
    \vspace{-5cm}
    \caption{Time of arrival at at $y=1$ of a particle that starts at $y=3$ with speed $v=-0.1$. Due to the small speed, interference between the ingoing and reflection at the potential becomes quite apparent.}
\end{figure}
The dispersion stemming from the standard approach not only yields arrival times in the forbidden region but also differs significantly from the probability distribution computed with our approach. For the same values of the parameters that describe an incoming particle with $v=-0.1$, the resulting distribution is shown in figure 4. Notice that the standard approach only provides non vanishing probabilities for values of $T>0$. The paradox arises because for negative $y$, in the forbidden region, it provides a non trivial distribution for $T$.

\begin{figure}[h]
    \centering
    \includegraphics[scale=0.45]{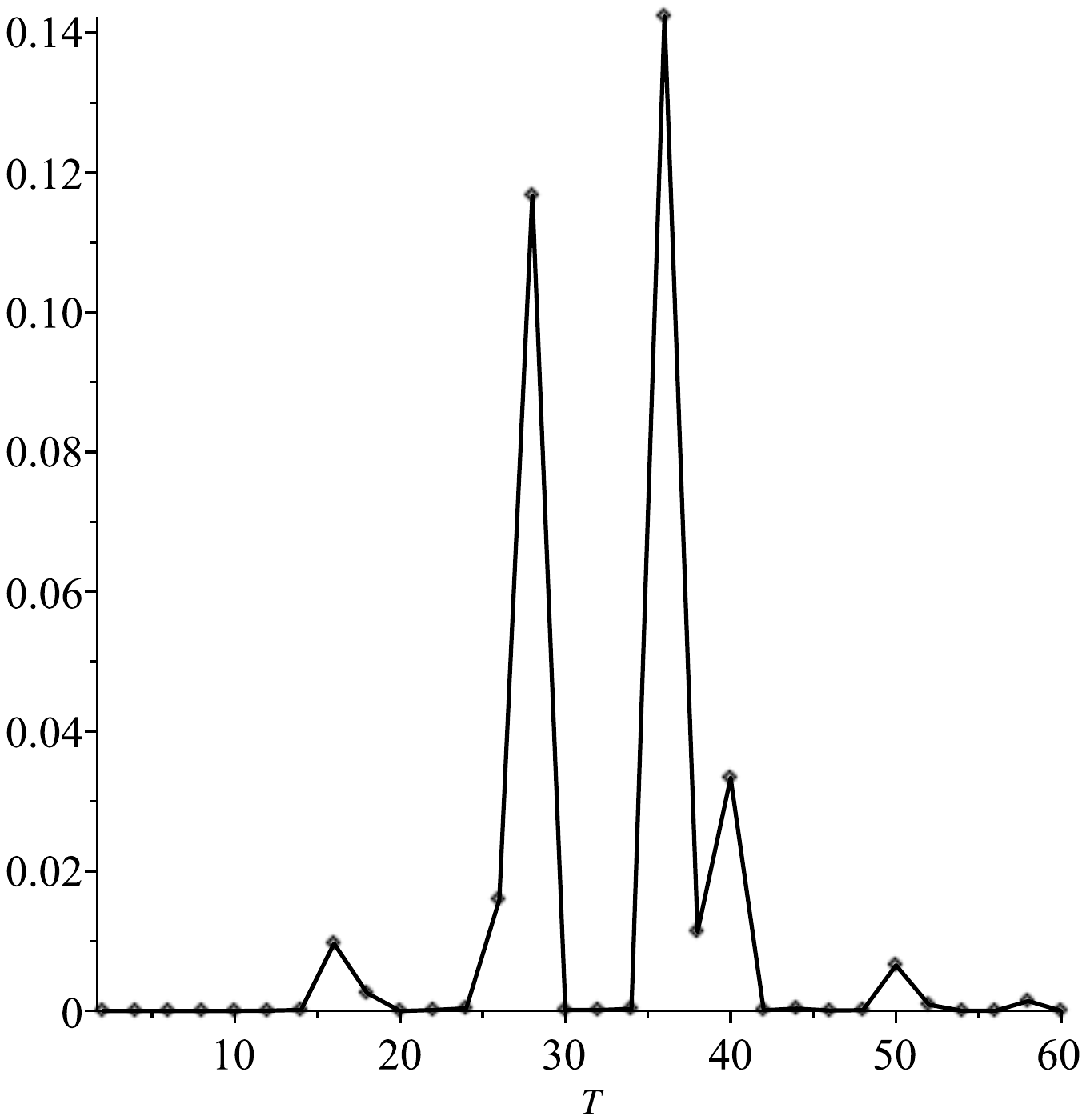}
    \vspace{-5cm}
    \caption{Time of arrival at $y=1$ for a packet prepared at $y=3$ in the case of an infinite potential barrier at $y=0$ computed in the standard approach. We only plot a limited number of points, as they are computationally costly. As can be seen, the predictions differ significantly from the approach presented in this paper. }
\end{figure}

Other paradoxical behaviors of the standard approach are eliminated by our proposal. Leavens \cite{Le} observed that a free particle in an antisymmetric superposition around $y=0$ has several anomalous behaviors. The wavefunction he obtains is,
\begin{equation}
    \psi^{(f)}(y,0)=N \left[
    \exp\left(-\frac{\left(y-y_0\right)^2}{4\left(\Delta y\right)^2}+i k_0 y\right)
    -\exp\left(-\frac{\left(y+y_0\right)^2}{4\left(\Delta y\right)^2}-i k_0 y\right)
    \right],\label{34}
\end{equation}
with,
\begin{equation}
    N=\left\{
    2^{3/2} \pi^{1/2} \Delta y 
    \left[ 
    1-
    \exp\left(-2\left(\Delta y\right)^2 k_0^2-\frac{y_0^2}{2 \left(\Delta y\right)^2}\right)
    \right]
    \right\}^{-1/2}.
\end{equation}
That is, a coherent linear superposition of Gaussians with centroids at $\pm y_0$, mean wavenumbers $\pm k_0$ and equal spatial widths $\Delta y$. Although the probability density $\vert \psi^{(f)}(y,t)\vert^2$ vanishes at $y=0$, the standard approach assigns finite probabilities to the times of arrival at $y=0$. Moreover, in spite that the evolution of the wavepacket $\vert \psi^{(f)}(y,t)\vert^2$ for $y>0$ is identical to the case of the infinite barrier, the distribution for the time of arrival is different in both cases. This is due to the Fourier transform of the wavefunctions for the infinite barrier $\psi(k,T)$ and for the superposition $\psi^{(f)}(k, T)$ are different and therefore $\Pi_\pm$ will be too. When one applies the approach we present in this paper both paradoxes disappear. The time of arrival to $x=0$ is not defined (it is $0/0$) and the probability distribution for arrival at the infinite barrier and the one resulting from the superpositions given in (\ref{34}) are identical. The result cannot be surprising since our distribution is computed from the probability density stemming from the wavefunction $\psi^{(f)}(y,t)=\psi_2(y,t)$ (with $\psi_2$ given by (\ref{26})) for $y>0$ without ever requiring a Fourier transform. 

\section{Time of arrival with real clocks}

We have discussed how to compute the time of arrival using conditional probabilities in the limit of an ideal clock. Physical systems evolve with Hamiltonians that are bounded-below. In particular, that would apply to the Hamiltonian of a real physical clock. The previous sections' analyses can be applied to real clocks. For instance, one can study a free particle using another free particle as a physical clock. Starting from equation (\ref{11}) with $\rho_1$ and $\rho_2$ states of the free particle for particles $1$ (clock) and $2$ respectively we have that the denominator of equation (\ref{11}) is given by,
\begin{equation}
    \int_{-\infty}^\infty dt {\rm Tr}\left(\rho(y) \rho(t)\right)=
    \int_{-\infty}^\infty dt 
    \sqrt{\frac{2}{\pi}}\,
        \sigma_y m_2 \chi_2^{-1/2}\exp\left(-
    \frac{2 \sigma_y^2 \left(\left(y-y_0\right)m_2-t p_2^{(0)}\right)^2}
    {\chi_2}
    \right) \Theta(t)
\end{equation}
with $y_0$ the initial position and $y$ the arrival point and $\chi_2=4 \sigma_y^4 m_2^2 + t^2$. We have assumed that the state was prepared at $t_0$ with a packet width $\sigma_y$ and $p_2^{(0)}$ is the initial momentum and $m_2$ the mass. 

On the other hand, the numerator of (\ref{11}) is given by,
\begin{eqnarray}
    &&\int_{-\infty}^\infty dt 
    {\frac{2}{\pi}}\,\chi_2^{-1/2}\sigma_y m_2 \chi_1^{-1/2}\sigma_x m_1 \nonumber\\
&&\times        \exp\left(-
    \frac{2 \sigma_y^2 \left(\left(y-y_0\right)m_2-t\, p_2^{(0)}\right)^2}
    {\chi_2}
    \right) 
        \exp\left(-
    \frac{2 \sigma_x^2 \left(\left(x-x_0\right)m_1-t\, p_1^{(0)}\right)^2}
    {\chi_1}
    \right) 
    \Theta(t)
\end{eqnarray}
with $\chi_1=4 \sigma_x^4 m_1^2 + t^2$ and $p_1^{(0)}$ the initial momentum of particle $1$ and $m_1$ its mass. 

Figure 5 shows the distribution of probabilities of arrival times. With the parameters chosen for the figure, the probability distribution of arrival times is not symmetric anymore, as is was in figure 1. Due to the dispersion of the physical clock it decreases more slowly than it grows as is shown in the figure. This is due to the widening of the clock's wavepacket as it propagates to larger values of $t$.
\begin{figure}[h]
    \centering
    \includegraphics[scale=0.45]{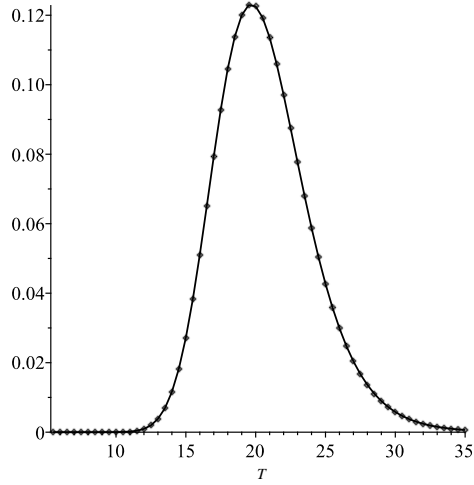}
    \vspace{-5cm}
    \caption{Probability of times of arrival with a real clock for the case of a particle at $y_0=3$, $y=1$, $m_2=200$, $p_2^{(0)}=-20$, $\sigma_y=0.2$, $v_2=-0.1$ and for the clock at $x_0=0$, $m_1=5000$, $p_1^{(0)}=5000$, $\sigma_x=0.02$. }
\end{figure}

\section{Time-energy uncertainty relation}

It is straightforward to show, within this framework, that time-energy uncertainty relations for the clock hold at the instant at which an event occurs,like the arrival of a particle to a given position. This can be done with the same techniques as in ordinary quantum mechanics one does for conjugate observables like $\hat{x}$ and $\hat{p}$. It is also possible to obtain analogous relations for arbitrary observables $\hat{A}$ and $\hat{B}$ of the clock  such that $[\hat{A},
\hat{B}]=i \hat{C}$  

Taking into account that any Dirac observable, with either discrete or continuous spectrum, can be written as,
\begin{equation}
    \hat{A} = \sum_i a_i \vert a_k\rangle\langle a_i\vert\quad{\rm or}\quad \int da \vert a\rangle\langle a\vert,
\end{equation}
equation (\ref{eq3}) allows to compute the expectation value of $\hat{A}$ (an observable associated with a clock) when $\hat{y}$ is observed as,
\begin{equation}
    \langle \hat{A}\rangle_y=
    \lim_{\tau\to\infty} 
    \frac{\int_{-\tau}^\tau dt \langle \hat{A}\rangle_t {\rm Tr}\left( P_y(t) \rho_2\right)}
    {\int_{-\tau}^\tau dt  {\rm Tr}\left( P_y(t) \rho_2\right)}.
\end{equation}

From this definition it is straightforward to derive uncertainty relations for the clock when an event, like the arrival of a particle to a point $y$, occurs. This is done  using standard techniques (see for instance \cite{cohen} page 286) and to show that if 
\begin{eqnarray}
\Delta A_y^2&\equiv& \left\langle \left(\hat{A}-\langle\hat{A}\rangle\right)_y^2\right\rangle_y,\\
\Delta B_y^2&\equiv& \left\langle \left(\hat{B}-\langle\hat{B}\rangle\right)_y^2\right\rangle_y,\\
\left[\hat{A},\hat{B}\right]&=&i \hat{C},
\end{eqnarray}
then,
\begin{equation}
    \Delta A_y \Delta B_y\ge \frac{1}{2}\langle\hat{C}\rangle_y.
\end{equation}

For the ideal clock with $\hat{H}=\hat{p}_1 c$ and therefore $[T,H]=i$ we have that $\Delta T \Delta E\ge 1/2$ and the usual expression is recovered. For real clocks the analysis is more delicate, and will be analyzed in detail in a separate paper.

\section{Conclusions}

We have applied the relational definition of time proposed as a solution to the problem of time in quantum gravity to the problem of time of arrival in quantum mechanics. We show that it yields satisfactory results, avoiding all of the problems and paradoxes that previous approaches had encountered. Since  each event gets assigned a projector the present solution of the time of arrival problem allows to assign a time to the occurrence of any event in a system. It has the advantage of being the only approach applicable in generally covariant systems and of treating the clock as any physical system.

\section*{Acknowledgements}
We thank Philipp H\"ohn and Luis Pedro García Pintos for comments.
This work was supported in part by Grant NSF-PHY-1903799, funds of the
Hearne Institute for Theoretical Physics, CCT-LSU, Pedeciba, and Fondo Clemente Estable FCE 1 2019 1 155865.

\end{document}